# Agile-CMMI Alignment: CMMI V2.0 Contributions and To-dos for Organizations

**Valeria Henriquez**
Universidad Politécnica de Madrid
Universidad Austral de Chile

**Ana M. Moreno**
Universidad Politécnica de Madrid

**Jose A. Calvo-Manzano**
Universidad Politécnica de Madrid

**Tomas San Feliu**
Universidad Politécnica de Madrid

*Abstract*—CMMI and Agile can work together. Over 80% of CMMI appraisals in 2018 were conducted at agile organizations, even though pre-2018 CMMI versions do not provide guidelines for agile contexts. A number of experience reports and research studies address the alignment between the two approaches but also pinpoint open tactical and organizational challenges. CMMI V2.0, published in 2018, was designed to be understandable, accessible, and flexible. It was intended to be integrated with other methodologies such as Agile. In this paper, we discuss to what extent the new CMMI V2.0 addresses the existing Agile-CMMI alignment challenges. We identify the two most significant CMMI V2.0 artifacts for this aim, the *context-specific sections* provided for most of the practice areas, and the *value statements* linked to the practices. We analyze how they contribute to each of the existing challenges and highlight important issues that organizations still need to tackle regarding this alignment.

*Keywords:* Agile and CMMI V2.0, CMMI V2.0, Scrum, Software Process Improvement and Agile Software Development

## 1 INTRODUCTION

CMMI is one of the most adopted performance improvement models worldwide [1]. CMMI practices indicate how organizations can successfully evolve their capability in particular areas to get business value. From its inception in 1987 (CMM) until the latest version published in 2018 (CMMI V2.0), the model has evolved to adapt to the changes taking place in industry.

On the other hand, agile practices are popular among many organizations. The Annual State of Agile Report published in 2020 states that 95% of surveyed organizations apply agile practices [2].

Although Agile and CMMI operate at orthogonal layers (CMMI is related to the "What" layer and Agile practices to the "How" layer) [3][4], both provide values, principles and best practices to pursue business goals and deliver value to the customer. Therefore, it is no surprise that the joint adoption of both approaches became popular. Over 80% of CMMI appraisals in 2018 were at agile organizations [5], even though the then CMMI V1.x did not provide specific guidelines for organizations adopting agile practices. This led to the publication of many industrial experiences and methodological papers analyzing the coexistence of CMMI V1.x and Agile (e.g., [6][7][8]). These papers set out their own proposals, defining associations between CMMI V1.x components and agile practices. These ad hoc proposals made a significant contribution to CMMI and Agile alignment, but also pinpointed several challenges for organizations, such as, limited recommendations about the specific agile practices or tools to be used, missing Agile guidelines for high maturity levels, or the risk of losses in terms of agility due to alignment [6][7][8].

CMMI V2.0, published in 2018, was designed to be understandable, flexible and applicable with approaches such as Agile [1]. This new model includes specific guidance to help organizations using agile practices (particularly Scrum) to strengthen their processes and scale up their agile practices with a focus on performance [1]. Therefore, the questions that arise are: Does CMMI V2.0 help to solve the existing Agile-CMMI alignment challenges? Which challenges, if any, do organizations still need to address?

As CMMI V2.0 has not been around for very long, there is hardly any published evidence to answer the above questions. To date, we have found only one publication describing a specific experience of alignment between CMMI V2.0 and the Scaled Agile Framework (SAFe) [9]. However, it does not fully answer the questions raised above.

In this work, we apply a three-step process to address Agile-CMMI V2.0 alignment:
1. We create a consolidated list of challenges facing organizations when aligning Agile and CMMI V1.x based on both a tertiary study of existing literature reviews and a secondary study.
2. We discuss which artifacts provided by CMMI V2.0 help to address the existing challenges, and how.
3. We identify a list of issues that software organizations still need to tackle as part of an Agile-CMMI V2.0 alignment process. Based on our experience, we present some pointers on how to address such issues, which, however, will depend on each individual software organization.

These results are particularly useful for senior managers to: 1) decide whether or not to commit to a process improvement effort; and 2) if so, properly specify the process improvement scope and resources allocation. Note that lack of commitment and support from higher-level management is one of the key reasons why process improvement initiatives fail [10].

## 2 AGILE-CMMI ALIGNMENT CHALLENGES

We followed a systematic mapping process (SMP) [11] to perform our tertiary study on already published CMMI V1.x and Agile literature. The research question (RQ) was: Which challenges concerning CMMI V1.x and Agile development have been identified over the last decade?

The string used to search the ACM, IEEE, ISI Web of Science, Science@Direct, Scopus and SpringerLink databases was: (Agile AND ("systematic review" OR "systematic literature review" OR "systematic mapping" OR "mapping study" OR "systematic map" OR "systematic mapping studies") AND ("maturity model" OR "maturity models" OR "CMMI")). The search period was from January 2010 to July 2019.

We used Parsifal (https://parsif.al) to remove duplicate studies, select studies according to the inclusion and exclusion criteria, and evaluate quality. The details of the tertiary study are available at https://bit.ly/2nO2J5E. The search returned 343 papers. After eliminating duplicates and irrelevant studies, we retrieved five secondary publications relevant to our research question. The details of these secondary papers are reported in the sidebar (references A1-A5). The five retrieved secondary studies cover a total of 230 primary papers (139 unique sources) published before 2016.

To the best of our knowledge, no secondary paper has been published later than 2016. To confirm that the challenges identified by pre-2016 publications are still unaddressed, and no new challenges have arisen, we conducted a secondary study (addressing the abovementioned RQ, data sources, and guidelines [11]) covering primary studies published from 2016 to 2019. The search string applied in this second SMP was: ((Agile OR Scrum OR XP) AND ("maturity model" OR "maturity models" OR CMMI)). As result we identified ten primary studies (see the sidebar, references B1-B10). The details of this second SMP are available at https://bit.ly/2PjQskH.

Validity threats to both the tertiary and secondary study were managed as described by [11]. To assure search coverage, we selected the best-known scientific databases. We also defined a search string with several keywords to improve the paper retrieval accuracy. To avoid study selection bias, we defined detailed inclusion and exclusion criteria, and all authors agreed upon the papers finally selected according to these criteria. Additionally, the search was rounded out by a snowballing approach. Finally, we dealt with the data extraction and classification threat by having all authors jointly analyze and evaluate the information gathered from the publications to ensure that the selected data were meaningful.

**Table 1** summarizes the identified alignment challenges. Depending on their scope, they were classified as tactical and organizational challenges. All the alignment challenges, except for *Standardization,* identified by the tertiary study (Table 1, top) were also detected by the primary studies (Table 1, bottom). As we will see later, Standardization addresses the alignment guideline comparability.

### 2.1 TACTICAL CHALLENGES

Organizations creating and/or applying existing agile alignment recommendations at project level face several difficulties related to the following aspects.

**Prescriptiveness**

Most secondary, and some primary, papers mention that existing alignment recommendations are descriptive and do not provide specific details on practices, techniques, and/or agile tools related to CMMI V1.x components. [A1] and [A3] also highlight that insufficient prescriptiveness is a major obstacle to determining complexity, costs, and risks, as well as to generating metrics and managing data, including privacy and security policies.

**Coverage**

According to [A1] and [A3], existing agile alignment recommendations do not cover all the CMMI V1.x components for all five maturity levels. In particular, the analyzed studies are based on agile practices described in Scrum and XP. Therefore, at best, they cover team/project management practices. Like the secondary studies, the primary papers mostly focus on Maturity Levels 2 and 3, and do not deal with the organizational issues covered by Maturity Levels 4 and 5. Only [B1], [B3], and [B6] refer to Maturity Levels 4 and 5, and propose preliminary, albeit non-generalizable, solutions.

**Standardization**

Each existing alignment recommendation presents associations between agile practices and CMMI V1.x components in different formats. As a result, they are not comparable. This is an obstacle to the selection of suitable guidance for the context of a particular organization [A3].

### 2.2 ORGANIZATIONAL CHALLENGES

A challenge perceived by organizations aligning Agile and CMMI V1.x is the project-level to organizational-level transition. In this transition, usually called cultural change [A2] [A3] [B2] [B4] [B8], organizations define and follow through their own work practices until they become part of the organizational culture. Based on the literature, challenges to this transition are related to the following aspects.

**Organizational guidelines**

Organizations aligning Agile and CMMI V1.x must define tailoring guidelines. Bearing in mind the range of projects executed by the organization, these guidelines should describe the contexts in which methods, techniques or practices related to the different processes enacted within the organization are to be used. Two of the secondary, and most of the primary, studies draw attention to the fact this is usually an uphill job because existing organizational processes are so ingrained in the company that they are difficult to question and change. [B1] also underscores that this is a tough undertaking unless the organization is open-minded enough to come up with new, goal-driven ways of performing organizational activities.

**Table 1. Alignment challenges reported by each secondary paper gathered from the tertiary study (2010 - 2019) and by the primary papers identified in the secondary study (2016 - 2019).**

|  |  | Tactical challenges | | | Organizational challenges | | |
|---|---|---|---|---|---|---|---|
|  |  | Prescriptive-ness | Coverage | Standardization | Organizational guidelines | Scaling | Agility |
| **Secondary papers** (covering 139 different primary papers) | A1 (24 primary papers) | X | X |  |  | X |  |
|  | A2 (34 primary papers) | X |  |  | X | X | X |
|  | A3 (81 primary papers) | X | X | X |  | X | X |
|  | A4 (52 primary papers) |  |  |  | X | X | X |
|  | A5 (39 primary papers) | X |  |  |  | X |  |
| **Primary papers** (10 papers) | B1 | X | X |  | X |  | X |
|  | B2 |  | X |  | X | X |  |
|  | B3 |  | X |  |  |  |  |
|  | B4 | X | X |  | X |  |  |
|  | B5 |  | X |  | X |  |  |
|  | B6 | X | X |  | X | X |  |
|  | B7 |  | X |  |  |  |  |
|  | B8 |  | X |  |  | X |  |
|  | B9 |  | X |  |  | X | X |
|  | B10 |  |  |  | X | X | X |

**Scaling experiences**

Organizations should systematically disseminate Agile and CMMI alignment experiences supported by their senior management. Otherwise, the experiences will be confined to teams/projects and not spread and adopted throughout the organization. This aspect is mentioned by all the secondary studies and some of the primary studies that address high maturity levels. Additionally, [B2] suggests that the mechanisms for disseminating the practices developed by specific teams/projects throughout the organization should also be agile.

**Agility**

Most secondary studies mention that CMMI V1.x can be aligned with agile techniques. However, they also underscore the risk of the loss of agility as organizations move up the CMMI V1.x maturity level scale. One exception ([B1]) states that "we found that our agile process actually worked more effectively after applying CMMI", although generally Agile and CMMI V1.x are regarded as opposite ends of the scale.

## 3 CMMI V2.0 MODEL

The CMMI V2.0 suite was published in 2018. According to [12], CMMI V2.0 "is an integrated set of best practices that enable businesses to improve performance of their key business processes".

The suite is composed of five products. Our paper focuses on the model because it details the best practices to improve performance [12]. Below, we run through this model in terms of its architecture, and its capability and maturity levels.

## 3.1 CMMI V2.0 MODEL ARCHITECTURE

The model architecture has the following components:
- *Categories for Capability Areas (CCAs)* are a logical grouping of capability areas.
- *Capability Area (CA)* is a set of practice areas that collectively describe a specific capability.
- *Practice Area (PA)* is a set of practices that describe a critical activity necessary to achieve a business goal and value.
- *Practice Group (PG)* is a logical grouping of practices at the same capability level. It provides an improvement roadmap for each PA.
- *Practice* is the most detailed level of the model. It describes the goal to be achieved and the business value that it generates.

CMMI 2.0 also provides *value statements* for each practice. They help to map business goals to CMMI components. **Figure 1** is a top-down illustration of the CMMI V2.0 architectural components (from CCA to Practice) for the Estimating (EST) PA, specifying the value that each practice generates [12].

Additionally, the PAs and practices might also include *context-specific* sections. These sections with a predefined structure contain guidelines clarifying the meaning of PAs or practices in particular contexts [12]. Some agile contexts mentioned by the CMMI Institute are "Agile with Scrum, DevOps, Kanban, Lean…" [12], although, to date, the CMMI-DEV only contains Agile with Scrum guidelines.

## 3.2 CMMI V2.0 CAPABILITY AND MATURITY LEVELS

Organizations adopting CMMI V2.0 can develop their improvement roadmap basing on *capability levels* or *maturity levels*. In both cases, there are five levels, ranging from Level 1 (Initial) to Level 5 (Optimizing), which build upon each other.

The *capability level* is well suited for organizations that want to improve in a particular area. For example, an organization that wants to accurately estimate project costs should evolve to Capability Level 2 of the EST PA. To do this, it must adopt Level 1 and 2 Practices for the Estimating (EST), Infrastructure (II), and Governance (GOV) PAs. Notice that no capability level can be achieved unless the respective II and GOV PAs are adopted.

On the other hand, *maturity levels* are well suited for organizations primarily performing a particular activity, for example, software development. In this case, CMMI V2.0 provides the CMMI-DEV view. This view describes a complete roadmap (with 19 PAs) for transitioning from ad hoc software development activities to disciplined and consistent processes to achieve business goals.

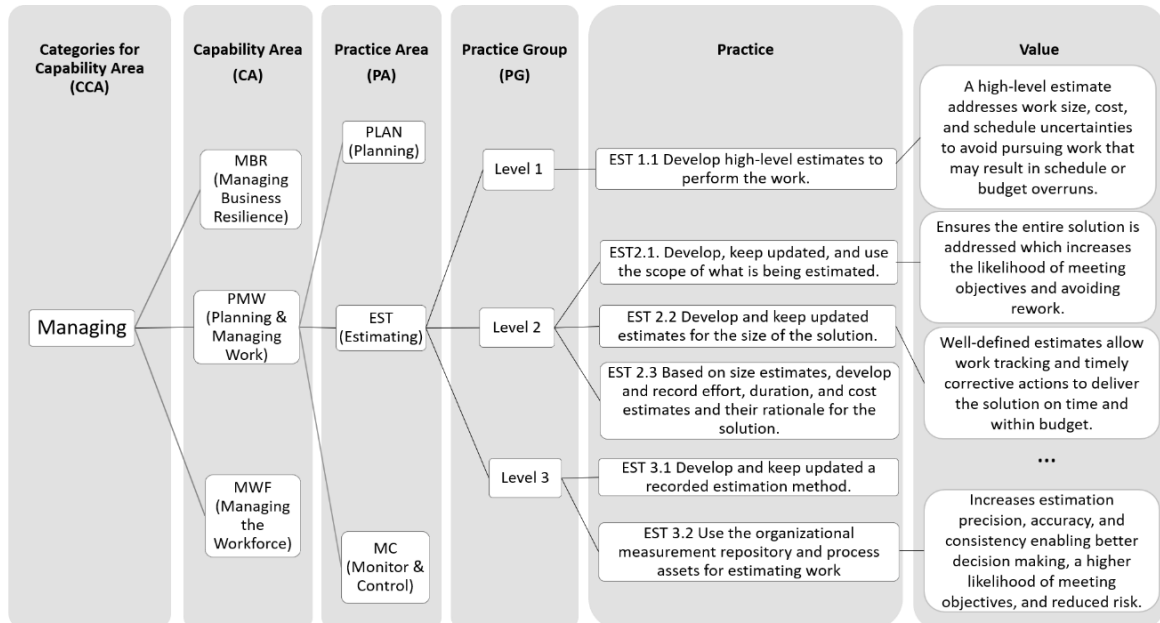

**Figure 1. Description of the Estimating PA (Capability Levels 1, 2 and 3) based on the CMMI V2.0 architecture.**

# 4 CMMI V2.0 CONTRIBUTIONS TO AGILE ALIGNMENT CHALLENGES

Through its *context-specific* sections and *value statements,* CMMI V2.0 makes a significant contribution to addressing Agile alignment challenges. On the context-specific side, even though only Agile with Scrum guidelines have been provided so far, the number, structure, and content of these sections contribute to Agile-CMMI alignment in different ways.

Regarding the *number of Agile context-specific* sections, CMMI-DEV V2.0 provides these sections in 16 out of its 19 PAs. However, it does not provide any *Agile context-specific* sections for practices. This means that three PAs and all the practices (for the 19 PAs) remain uncovered. Even so, CMMI V2.0 has made a major effort to address the **coverage challenge**, providing guidelines to align Scrum to most of the CMMI-DEV PAs.

Moreover, some of these *Agile context-specific* sections are provided for PAs that deal with organizational issues. These PAs are: Managing Performance and Measurement (MPM); Process Asset Development (PAD); and Process Management (PCM). They help to define processes, assets, and metrics to repeat successful performance throughout the organization [12]. Therefore, even though agile guidelines are provided only for the previous PAs and not for their practices, CMMI V2.0 helps to reduce the challenges of **creating organizational guidelines and scaling experiences**.

On top of the agile guidelines for the organizational PAs, the *value statements* provided for CMMI-DEV practices set an explicit business goal to be achieved and help to find agile techniques and tools to generate such goals properly. In this regard, *value statements* help to address the creation of meaningful and goal-driven guidelines, which is key to resolving the **organizational guideline challenge** [B1]. In the same vein, the *value statements* for the GOV and II PAs explicitly focus on the business value generated by engaging the senior management in process implementation and improvement. This information is useful for tackling the root cause of the **scaling challenge***,* namely, the lack of senior managers promoting organizational change [A2]. *Value statements* help senior managers to better understand that their participation in scaling initiatives minimizes the process implementation cost, ensures organizational alignment, and increases the likelihood of achieving business objectives [12].

Regarding the *structure of the context-specific* sections, CMMI V2.0 provides a unique format to create Agile-CMMI guidelines. In particular, it suggests three fields: identifier, description, and explanation. If needed, other fields, like value, intent, example activities, work product examples, and external links can be added. Thus, this structure definition is a major contribution to resolving the **standardization challenge**. The use of this standard format could facilitate both the spread of knowledge throughout the organization and knowledge sharing among organizations.

The *content of the Agile context-specific* sections differs depending on the PA. Some PAs have more and/or longer explanations than others. In line with CMMI philosophy, they describe what to do, but not how to do it [12]. CMMI V2.0 is not a set of implementable processes, and its Agile guidelines are purposely not prescriptive. However, the *Agile context-specific* sections of most of the PAs related to CMMI-DEV Maturity Level 2 include specific details that can contribute to satisfy the demand for prescriptiveness referred to in the literature. For instance, the Agile guideline for EST shows which steps are required to perform estimations in Scrum events and provides examples of the information to be included in the backlog records to adopt the EST PA. Additionally, the *value statement* in each CMMI-DEV practice can be used by organizations to check if their prescriptive guidelines truly help to achieve their business goals. In this sense, *value statements* also contribute, indirectly, to resolving the **prescriptiveness challenge**.

Finally, the *content of the Agile context-specific* sections refers to the Scrum event in which a PA generally takes place, or how the adoption of CMMI-DEV V2.0 practices can strengthen Scrum practices. However, there are no specific recommendations for **maintaining agility** during the adaptation of such agile practices. Therefore, CMMI-DEV V2.0 does not explicitly mention how to deal with this risk. For example, the Agile guideline for the Requirements Development and Management (RDM) PA mentions that Scrum teams should modify standard sprint execution to tackle aspects like constraints, interfaces or connections, and quality attributes. These aspects can add important business value to some software projects, but could, at the same time, also have a major impact on agility.

# 5 CMMI V2.0 AND AGILE ALIGNMENT: TO-DOS FOR ORGANIZATIONS

As we have seen, CMMI V2.0 helps to address most of the existing Agile-CMMI alignment challenges, describing how to align most of the PAs with Scrum, and assisting organizations to ensure that their agile practices are aligned with business value provision in the context of CMMI V2.0. However, there are still important to-dos for organizations aligning Agile and CMMI V2.0, either from scratch or as a transition from CMMI V1.3.

Below, we detail these to-dos and provide some pointers to guide organizations through this alignment. These pointers are based on both the literature and our two-decade-long software process improvement consulting experience. However, we aim to highlight the Agile-CMMI V2.0 alignment challenges in the knowledge that there are no one-size-fits-all solutions. Ultimately, each organization must decide how to implement its improvement initiatives in its unique context.

- *Contextualize GOV, II, and OT to Agile*—The CMMI V2.0 highlights the importance of senior management involvement, defining the GOV and II PAs as a mandatory part of any roadmap. However, CMMI V2.0 does not provide Agile guidelines for these PAs. Organizations could explore the possibility of providing such contextualizations based on SAFe [13], or alternatively DAD [14], which are Agile frameworks that address organizational issues [15]. Senior managers may explore these Agile approaches to figure out how to govern (GOV), allocate infrastructure (II), and train (OT) in an agile way to scale up processes that ensure repeatable business success. Not only should they encourage the adoption of agile values, like transparency, collaboration, fail-fast culture, and self-organization, as organizational guidelines, they should also lead by example [4][13][14] [16][17].
- *Develop Agile contextualizations at practice level*—Agile coverage at PA level is high (84%) in CMMI-DEV V2.0. However, organizations are responsible for sizing the effort required to agilely implement the CMMI V2.0 practices that help to achieve their own business goals. To do this, one possible line of action is to map the Agile guidelines for CMMI-DEV V1.3 [3] to CMMI-DEV V2.0 practices. This Agile guide [3] details how 20 agile practices (described in Scrum and XP) can be used in conjunction with some CMMI-DEV V1.3 components. The CMMI V2.0 architecture is not directly equivalent to the CMMI V1.3 model. However, there is an official correspondence for transitioning between model versions [18]. Although these two documents can help to develop agile contextualizations at practice level, this task requires considerable effort.
- *Develop prescriptive guidelines*—CMMI V2.0 is not a one-size-fits-all solution. In fact, it is up to each individual organization to choose agile methods, techniques, and tools to perform its activities in its unique context. Senior managers should earmark resources for developing prescriptive guidelines for the PAs that are most related to their business goals. As already mentioned, Maturity Level 2 PAs have the most detailed Agile context-specific sections and they can serve as inspiration to round out other CMMI PAs.
- *Control the impact on agility*—Cross-cutting the above points, particular care should be taken to constantly monitor agility. Although there is general agreement that it is possible to align CMMI and Agile [A1-A5], there is no clear evidence regarding the impact of this alignment on agility and, ultimately, adaptability to market needs. As a result, it is essential for senior managers to ensure data are collected and analyzed to monitor agility during alignment with CMMI V2.0. Depending on variables like software type, quality criteria, and user idiosyncrasy, specific metrics and thresholds should be defined to continuously control agility indicators for the organization or even project [13][19]. On the other hand, the above Agile organizational frameworks might also help to address this task with a set of activities for assessing agility (e.g., see the *Measure and Grow* section in [13]).

# 6 DISCUSSION

**Table 2** summarizes the results of this research. For each Agile-CMMI alignment challenge identified from the literature (Table 2 rows), the first column lists the CMMI V2.0 contributions, while the second column presents the open issues that Agile organizations still need to address.

Below we discuss the threats to the validity of our results using Maxwell's categorization [20].

Overall, we recognize a potential bias towards positive outcomes (interpretive validity threat), as we support the idea that CMMI and Agile can be effectively aligned.

A closer look at the findings concerning the contribution of CMMI V2.0 to the alignment challenges reveals that such results are primarily descriptive. Therefore, their accuracy or descriptive validity can be checked against the content of the CMMI V2.0 model. Note, however, that, during this description process, we made some decisions that deserve consideration. One such decision concerned how to measure coverage. We counted the number of context-specific sections labeled as "Agile with Scrum Guidance", which we compared with the total number of PAs for each CMMI-DEV V2.0 maturity level. We omitted the content of these sections, which was studied under another challenge. On the other hand, CMMI-DEV V2.0 does not explicitly provide any agile guideline for its practices. It could be argued, however, that a PA agile guideline also covers its practices. We are opposed to this idea because CMMI V2.0 does account for practice-level guidelines for other contexts, such as CMMI-SVC.

Regarding the to-dos for organizations (Table 2, Column 2), we aim, as already mentioned, to highlight the open issues that need to be addressed by organizations aligning Agile and CMMI V2.0. The reliability of these results also relates to descriptive validity because the to-dos can be seen as the complementary part of the CMMI V2.0 contributions. Additionally, we provide some pointers to address such to-dos based on evidence gathered from over 20 years' experience in the software industry as process improvement researchers and consultants (working on CMMI appraisals, and agile transformation). Even so, we recognize that this could have a potential impact on the external, or generalizability, validity of such recommendations. For example, our recommendation regarding standardization using the structure of the context-specific section suggested by CMMI V.2.0 is just one option. Organizations are at liberty to define their own agile guideline structure, but, if they do, it could be hard to reach cross-organizational agreement, and this would perpetuate the standardization challenge.

The external validity of the recommendations related to organizational guidelines, and scaling agility also warrant further discussion. In these cases, we have compared our experience and beliefs with the literature. The literature confirms that only senior managers have the authority to create an organizational environment that encourages improvement culture [4][13][19]. To do this, the leadership style must match the promoted change, that is, leaders must figure out how to perform their roles applying the practices that they promote [4][13][16][17]. Therefore, we suggest that senior managers seek inspiration from organizational agile frameworks, such as DAD, and, particularly SAFe [13], which is becoming increasingly popular [2]. But alternatively, an organization can also handle this issue using an ad hoc strategy [4][16][17].

Regarding the external validity of the recommendations on agility maintenance, we suggest monitoring indicators that somehow represent agility. We provide examples of some of the dimensions, also supported by [13] and [19], that an organization could consider when developing such agility indicators. However, as mentioned above, there is more than one approach to this issue, and organizations may each design an ad hoc configuration of their agility-related dimensions.

Note, finally, that organizations should thoroughly understand that the primary aim of aligning Agile and CMMI V2.0 practices is to repeat successful performance and achieve business goals following agile practices. Advance knowledge of the Agile-CMMI V2.0 alignment challenges will help organizations to truly commit to the improvement initiative, as well as allocate resources to better define a path forward towards Agile and CMMI alignment.

**Table 2. Contributions of CMMI V2.0 and to-dos for organizations aligning Agile and CMMI V2.0.**

|  |  | **CMMI V2.0 model contributions** | **CMMI V2.0 and Agile alignment to-dos and pointers for organizations** |
|---|---|---|---|
| **Tactical challenge** | **Prescriptiveness** | Guidelines on how to align Agile with Scrum work activities and products to get the business value related to some of the model PAs. Most of CMMI-DEV Maturity Level 2 PAs, such as Planning (PLAN), Estimating (EST) Configuration Management (CM), and Monitor & Control (MC), provide detailed examples of this alignment.<br>The value statement of each practice is useful for checking that the prescriptive alignment guidelines truly help to achieve business goals. | Senior managers of each organization should support the prescription of the agile methods, techniques, and tools needed to achieve their organizational objectives in their unique context.<br>The Planning (PLAN), Estimating (EST) Configuration Management (CM), and Monitor & Control (MC) PA guidelines can be a source of inspiration for rounding out other CMMI PAs. |
| | **Coverage** | Partial coverage of CMMI-DEV PAs with contextualization for Agile with Scrum in:<br><br>• 78% of Maturity Level 2 PAs (7 out of 9 PAs).<br>• 84% of Maturity Levels 3, 4 and 5 PAs (16 out of 19 PAs).<br><br>Notice that each maturity level is built upon each other. | CMMI-DEV V2.0 does not provide Agile coverage for the following PAs:<br>Organizational training (OT), Governance (GOV)*, and Implementation infrastructure (II)*.<br>If organizations decide to adopt any of the above PAs, then the senior manager should be actively involved and provide the delivery teams with support to contextualize the respective PAs with Agile.<br>*PA required to achieve any maturity/capability level in CMMI V2.0. |
| | **Standardization** | Definition of the context-specific section and its structure:<br><br>• Identifier and description<br>• Explanation<br>• Intent and value<br>• Example activities<br>• Context-specific work product examples | Senior managers should support the creation of Agile and CMMI alignment guidelines for their organization.<br>One option is to follow the context-specific section structure provided by CMMI V2.0. |
| **Organizational challenge** | **Organizational guidelines & scaling experiences** | Contextualization for Agile with Scrum in 50% of the PAs related to organizational matters (MPM, PAD, and PCM with Agile contextualization; GOV, II, and OT without Agile contextualization).<br>The value statements for organizational PAs lead the development of meaningful and goal-driven organizational guidelines.<br>The value statements of GOV and II PAs provide well-founded reasons to convince senior management to participate in scaling Agile-CMMI initiatives. | Senior management should develop agile guidelines on management role performance in pursuit of the GOV, II, and OT values. They should also support the development of organizational agile guidelines for MPM, PAD, and PCM practices.<br>Senior managers might refer to agile organizational frameworks to align GOV, II, and OT PAs to scale up processes and ensure repeatable business success. Senior managers should lead by example and adopt agile values as part of their own roles [13].<br>Some examples of agile strategies that senior management can adopt are: organize work around business value using value stream mapping; or decentralizing decision making based on frequency, cost of delay and information needs [4][13]. |
| | **Agility** | No explicit issues are identified. | Senior managers should ensure that agility is measured during alignment with CMMI V2.0.<br>Senior managers should address this task defining specific activities, metrics, and thresholds according to each organization or even project needs. See, for example, suggestions provided in [13] [19]. |

# 7 ACKNOWLEDGMENT

This work was partly supported by the CONICYT PFCHA/DOCTORADO BECAS CHILE/2018 - 72190378.

**Valeria Henriquez** is co-founder and CEO of a Chilean company. She also is a PhD student at the Universidad Politécnica de Madrid. Her research interests include agile methods in software engineering and software process improvement. Contact her at valeria.henriquez@puentesdigitales.cl.

**Ana M. Moreno** is full professor of Software Engineering at the Universidad Politécnica de Madrid. She received her PhD in Computer Science from the Universidad Politécnica de Madrid. Her research interests include software project management, software usability and agile methods. She is member of IEEE CS. Contact her at ammoreno@fi.upm.es.

**Jose A. Calvo-Manzano** is associate professor at the Universidad Politécnica de Madrid. He received his PhD in Computer Science from the Universidad de Vigo. His research interests include software process improvement focused on CMMI, software management processes and cybersecurity risks. Contact him at joseantonio.calvomanzano@upm.es.

**Tomas San Feliu** is associate professor at Universidad Politécnica de Madrid. He received his PhD in Computer Science from the Universidad de Castilla-La Mancha. His research interests include risk management, process improvement, and software metrics. Contact him at tomas.sanfeliu@upm.es.

**Sidebar**

| | Secondary papers gathered from the tertiary study (2010 - 2019) | | |
|---|---|---|---|
| **ID** | **Reference** | **Number of primary studies** | **Search period** |
| [A1] | C. J. Torrecilla-Salinas, J. Sedeño, M. J. Escalona, and M. Mejías, "Agile, Web Engineering and Capability Maturity Model Integration: A systematic literature review," Information and Software Technology, vol 71, pp. 92-107, 2016. | 24 | 2001 - 2015 |
| [A2] | L. F. Chagas, D. D. de Carvalho, A. M. Lima, C. Alessandra, and C.A.L. Reis, "Systematic Literature Review on the Characteristics of Agile Project Management in the Context of Maturity Models," in International Conference on Software Process Improvement and Capability Determination, 2014, pp. 177–189. | 34 | 2001-2013 |
| [A3] | F. Selleri et al., "Using CMMI together with agile software development: A systematic review," Information and Software Technology, vol. 58, pp. 20–43, 2015. | 81 | Until 2011 |
| [A4] | M. Palomino, A. Dávila, K. Melendez, and M. Pessoa, "Agile Practices Adoption in CMMI Organizations: A Systematic Literature Review," in International Conference on Software Process Improvement, 2017, pp. 57–67. | 52 | Until 2016 |
| [A5] | V. Henriques and M. Tanner, "A Systematic Literature Review of Agile and Maturity Model Research," Interdisciplinary Journal of Information, Knowledge, and Management, vol. 12, pp. 53–73, 2017. | 39 | 2000 – 2015 |
| **Primary papers identified from the secondary study (2016 - 2019)** | | | |
| **ID** | **Reference** | | |
| [B1] | D. Sharma, N. Narula, D. Lee, and T. R. Leishman, "Agile 5 using high maturity CMMI practices to improve agile processes and achieve predictable results," CrossTalk, vol. 29, no. 4, pp. 32–35, 2016. | | |
| [B2] | M. De Angelis and R. Bizzoni, "Agile Methodology in Progesi MDA Model (Meta–Dynamic–Agile)," in Proceedings of 4th International Conference in Software Engineering for Defence Applications, 2016, pp. 243-260. | | |
| [B3] | C. J. Torrecilla-Salinas, J. Sedeño, J. Escalona, and M. Mejías, "An Agile approach to CMMI-DEV levels 4 and 5 in Web development," in Information Systems Development: Complexity in Information Systems Development, 2016, pp. 137–149. | | |
| [B4] | D. Carlson and E. Soukup, "Combining Agile Practices with CMMI Process Areas A Case Study," CrossTalk, vol. 30, no. 4, pp. 17–21, 2017. | | |
| [B5] | A. B. Farid, A. S. Elghany, and Y. M. Helmy, "Implementing Project Management Category Process Areas of CMMI Version 1.3 Using SCRUM Practices, and Assets," International Journal of Advanced Computer Science and Applications, vol. 7, no. 2, pp. 243–252, 2016. | | |
| [B6] | C. J. Torrecilla-Salinas, T. Guardia, O. De Troyer, M. Mejías, and J. Sede, "NDT-Agile: An Agile, CMMI-Compatible Framework for Web Engineering," in International Conference on Software Process Improvement and Capability Determination, 2017, pp. 3–16. | | |
| [B7] | Z. Bougroun, A. Zeaaraoui, and T. Bouchentouf, "Scrumban / XP: A New Approach to Cover the Third Level of CMMI Model," in Proceedings of the Mediterranean Conference on Information & Communication Technologies, 2016, pp. 383–391. | | |
| [B8] | S. Kawamoto and J. R. De Almeida, "Scrum-DR: An extension of the SCRUM framework adherent to the capability maturity model using design rationale techniques," in CHILEAN Conference on Electrical, Electronics Engineering, Information and Communication Technologies (CHILECON), 2017, pp. 1–7. | | |
| [B9] | S. K. Amer, N. Badr, and A. Hamad, "Combining CMMI Specific Practices with Scrum Model to Address Shortcomings in Process Maturity," in International Conference on Advanced Machine Learning Technologies and Applications, 2020, vol. 921, pp. 898–907. | | |
| [B10] | R. Albuquerque, R. Fontana, A. Malucelli, and S. Reinehr, "Agile Methods and Maturity Models Assessments: What's Next?" in European Conference on Software Process Improvement, 2019, vol. 1060, pp. 619–630. | | |